\documentclass[twoside,twocolumn,english,pra,twocolumn,aps,superscriptaddress,showpacs]{revtex4}

\usepackage[T1]{fontenc}
\usepackage[latin9]{inputenc}
\usepackage{color}
\usepackage{babel}
\usepackage{textcomp}
\usepackage{amsmath}
\usepackage{amsthm}
\usepackage{graphicx}
\usepackage[unicode=true,pdfusetitle,
 bookmarks=true,bookmarksnumbered=false,bookmarksopen=false,
 breaklinks=false,pdfborder={0 0 0},backref=false,colorlinks=true]
 {hyperref}
\usepackage{breakurl}
\usepackage{epstopdf}
\usepackage{amsmath,bm}

\newcommand{\ehbar}{\hbar_{\mathrm{eff}}}

\begin{document}

\title{Effective protection of quantum coherence by non-Hermitian driving}

\author{Kai-Qian Huang}
\affiliation{
School of Science, Jiangxi University of Science and Technology, Ganzhou 341000, China}
\affiliation{
Guangdong Provincial Key Laboratory of Quantum Engineering and Quantum Materials, SPTE,South China Normal University,Guangzhou 510006,China
}
\author{Wen-Lei Zhao}
\email[]{wlzhao@jxust.edu.cn}
\affiliation{
School of Science, Jiangxi University of Science and Technology, Ganzhou 341000, China}

\author{Zhi Li}
\email[]{lizhiphys@126.com}
\affiliation{
Guangdong Provincial Key Laboratory of Quantum Engineering and Quantum Materials, SPTE, South China Normal University, Guangzhou 510006,China
}

\begin{abstract}
We investigate the effects of non-Hermitian driving on quantum coherence in a bipartite system. The results that the dynamical localization destroyed by the Hermitian interaction revives are an evidence of the restoration of quantum coherence by non-Hermitian driving. Besides, the entanglement between the two subsystems also decays with the boosting of non-hermitian driving strength, which provides another evidence that non-Hermitian driving will protect quantum coherence. The physics behind this phenomenon is the domination of the quasieigenstate with maximum imaginary value of the quasieigenvalue on the dynamics of the non-Hermitian system. Our discovery establishes a restoration mechanism of quantum coherence in interacting and dissipative quantum systems, which is highly relevant to experiments in diverse fields from many-body physics to quantum information.
\end{abstract}
\date{\today}
\maketitle

\section{Introduction}
As a fingerprint of quantum coherence, localization of particles is a longstanding problem in different fields of physics, such as condensate matter physics and quantum chaos. A landmark study of Anderson discovered that the diffusion of electronics in disordered potential is frozen, which is the well-known Anderson localization (AL)~\cite{Anderson58}. The phenomenon of AL has been observed in acoustic wave~\cite{Condat87a,Condat87b}, optics~\cite{Chabanov00,Segev13} and matter waves~\cite{Billy08,Roati08}. Thus intrinsically it is a universal behavior caused by wave interference. In periodically driven systems~\cite{Casati79,Izrailev90}, an analog of AL is the dynamical localization (DL) taking place in momentum space. Interestingly, the eigenequation of the Floquet operator of delta-kick systems is mathematically equivalent to that of the Anderson model, which makes the DL be on the same footing as AL, namely, both of them are the common phenomenon resulting from quantum coherence~\cite{Fishman82}. Remarkably, experimental progress has realized the DL of cold atoms subjected to periodical kicking by laser~\cite{Moore95,Ammann98,dArcy01,Ryu06,Talukdar10,Lopez12,Gadway13}, which greatly facilitates the investigation in the effects of inter-atomic interaction on quantum coherence and thereby leads to the appearance of novel transport phenomenon of matter waves~\cite{Zhaohong20}.

The behaviors of quantum coherence in the presence of inter-atomic interaction is a fundamental problem which has potential application in the fields of quantum information and quantum computation. Previous investigations in quantum chaos reported that the quantum coherence is fragile. As it is destroyed by sufficiently strong interaction, a wide spectrum of diffusion from the power-law diffusion $\propto t^{\alpha}$~\cite{Fishman12,Flach13,Veksler09,Pikovsky11,Michaely12,Vakulchyk18,Cherroret14,Gligoric13} to exponential diffusion $\propto e^{\beta t}$~\cite{Mieck05,Guarneri17,Zhao16,Zhao19JPA} has been found.
More recently, a theoretical study proves that an arbitrarily weak perturbation is enough to destroy AL, which demonstrates the disappearance of quantum coherence and provides a solid theoretical foundation for future investigations~\cite{Zhaohong20}. However, the issue of quantum coherence under many-body interaction is still inconclusive. Recent works report the existence of many-body localization, which indicates the appearance of quantum coherence~\cite{Rylands20}. On the other hand, the effects of interaction on quantum coherence is closely related to the fundamental problem of quantum-classical transition. It is known  that the interaction with an environment even consisting of one degree of freedom is able to destroy quantum coherence, and consequently causes the emergence of the classically-chaotic diffusion from the underlying quantum dynamics, namely, the appearance of quantum-classical correspondence (QCC)~\cite{rossi,bandy,adach,graha,park,Zhao09,Zhao10,Zhao20,Gadway,Rozenbaum17}.

The common understanding of the quantum decoherence by interaction is based on the assumption of Hermiticity of quantum mechanics. The rich physics in non-Hermitian systems has recently attracted extensive interests, since these intriguing behaviors in open quantum systems (e.g., optics propagation in dissipative medium, elementary excitation in condensate matter, the evolution of cold atoms with limited life-span, etc.) can all be described by non-Hermitian Hamiltonian. Exotic transport behaviors, for instance unidirectional reflection and nonreciprocal diffraction of optics, greatly promote the exciting application of non-Hermitian physics in the actual control of light~\cite{Suchkov16} and heat transport~\cite{Liying19}. In addition, non-Hermiticity has been recognized as a fundamental modification of quantum mechanics~\cite{Ashida20,Moiseyev2011}, which opens a new prospect for the theoretical physics. In this context, the novel physics of quantum coherence in the presence of interaction in non-Hermitian systems deserves urgent investigation.

In the present work, we investigate the effects of non-Hermiticity on the quantum coherence, which is displayed by the dynamics of quantum diffusion and entanglement via two coupled kicked rotors with non-Hermitian kicking potential.
Interestingly, the results reveal that strong enough non-Hermitian driving can destroy the QCC of the diffusion dynamics and lead to the appearance of DL, which is a signature of the recovery of quantum coherence~\cite{Zhao20b}. The entanglement of the subsystems is gradually reduced by increasing the strength of the non-Hermitian driving, even if the initial state is of maximum entanglement. The underlying physics is that an arbitrary state evolves to the quasieigenstate (QES) with maximum imaginary part of the complex quasieigenvalue (QEV). Our result clearly demonstrates the protection and recovery of quantum coherence by the non-Hermitian driving, which has potential application in the quantum information technology.

The paper is organized as follows. In Sec.~\ref{MResl}, we describe our model and show the phenomenon of dynamical localization induced by non-Hermitian driving. In Sec.~\ref{THEAnal}, we show our theoretical analysis. Summary is presented in Sec.~\ref{ConcL}.

\section{Model and Results}\label{MResl}
The Hamiltonian of the bipartite system reads
\begin{equation}\label{Hamil}
{\rm H} = {\rm H}_1 + {\rm H}_2 + {\rm H}_{\rm I}\;,
\end{equation}
where ${\rm H}_j$ ($j=1, 2$) represents the Hamiltonian of individual particles, and ${\rm H}_{\rm I}$ is the interaction. Each subsystem is a kicked rotor  for which the Hamiltonian takes the form
\begin{equation}\label{Hamil}
{\rm H}_j=\frac{{p_j}^2}{2}+ V_K(\theta_j)\sum_n
\delta(t-t_n)\:,
\end{equation}
with
\begin{equation}
V_K(\theta_j)= (K_j +{i} \lambda_j)\cos (\theta_j)\;,
\end{equation}
where $p_j$ is the angular momentum operator, $\theta_{j}$ is the angle coordinate, $K_{j}$ denotes the strength of the real part of the kicking potential, $\lambda_{j}$ is the strength of its imaginary part.
The interaction Hamiltonian has the expression
\begin{equation}\label{INTACTHamil}
{\rm H_I}=\varepsilon \ehbar \cos(\theta_1)\cos(\theta_2)\sum_n\delta(t-t_n)\;,
\end{equation}
with $\varepsilon$ being the interaction strength and $\ehbar$ the effective Plank constant. All quantities are in dimensionless units.
An arbitrary state can be expanded on the basis of the product states of the eigenstates of unperturbed rotors, i.e., $|\psi\rangle =\sum_{m,n}\psi_{m,n}|m,n\rangle$ with $p|n\rangle=n\ehbar |n\rangle$ and $\langle \theta|n\rangle = e^{in\theta}/\sqrt{2\pi}$.
The time evolution from $t=t_n$ to $t=t_{n+1}$ is governed by the Floquet operator $|\psi(t_{n+1})\rangle = U|\psi(t_n)\rangle$. Due to the delta kick, the Floquet operator can be separated into two fractions
\begin{equation}\label{FQTOpr}
U=U_f U_K\;,
\end{equation}
where the  evolution operator of the kinetic term takes the form
\begin{equation}
U_f=\exp\left( -\frac{i}{\ehbar}\sum_{j=1}^{2} \frac{p_j^{2}}{2} \right) \;,
\end{equation}
and for the kick term it is
\begin{equation}
U_K=\exp\left[{-\frac{i}{\ehbar}\sum_{j=1}^{2}  V_K(\theta_j)} -\frac{i}{\ehbar} {\rm H_I}(\theta_1,\theta_2) \right]\;.
\end{equation}

To quantify the quantum diffusion, we numerically calculate the mean square of momentum of one of particles (say particle 1)
\begin{equation}\label{MSMomentum}
\langle p_1^2 \rangle ={\rm Tr}(\rho_1 p_1^2)\;,
\end{equation}
with
\begin{equation}\label{MSMomentum}
\rho_1 =\frac{1}{\cal{N}}{\rm Tr}_2(|\psi\rangle \langle \psi|)\;,
\end{equation}
where $\rho_1$ is the reduced density matrix by using the partial trace to the density matrix of the two-particle system $\rho=|\psi\rangle \langle \psi|$, and $\cal{N}$ is the norm of quantum states. Note that the value of $\cal{N}$ will exponentially increase due to the appearance of complex QEVs for sufficiently strong non-Hermitian kicking. Therefore, the above definition of expectation value $\langle p_1^2 \rangle$ reduces the contribution of the growth from the norm. In numerical simulation, we set the initial state as the product state of the ground state $|\psi(t_0)\rangle = |0,0\rangle$. Without loss of generality, we consider the case with $K_1=K_2=K$ and $\lambda_1=\lambda_2=\lambda$, namely, the two particles are identical. We also investigate the case with $\lambda_1=0$ and $\lambda_2\neq0$, which mimics the situation where a Hermitian particle under the effect of non-Hermitian environment. And we find that there are no essential differences with $\lambda_1=\lambda_2$.

\begin{figure}[htbp]
\begin{center}
\includegraphics[width=8.5cm]{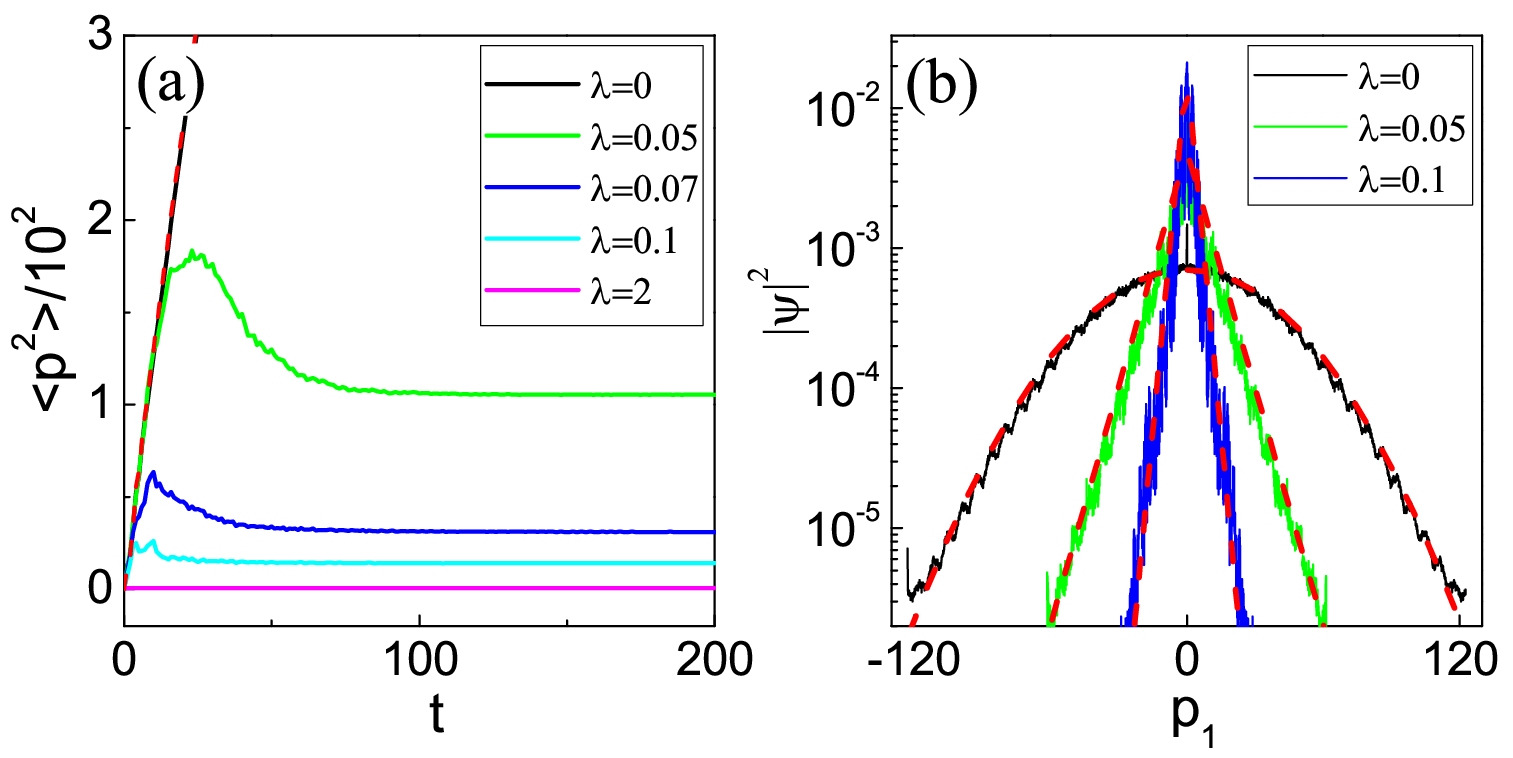}
\caption{(Color online) (a): Time dependence of the mean energy for particle 1 $\langle p^2_1\rangle$. From top to bottom, solid lines correspond to $\lambda=0$ (black), $0.05$ (green), $0.07$ (blue), $0.1$ (cyan) and $2.0$ (magenta). Red-dashed line  denotes the classical mean energy $\langle p^2_1\rangle$ with $\lambda=0$. (b): Probability density distribution in momentum space for particle 1 $|\psi_1|^2$ at the time $t_n=100$ with $\lambda=0$ (black line), $0.05$ (green line) and $=0.1$ (blue line).
Dash-dotted line (in red) denotes the Gaussian function of the form $|\psi(p_1)|^2 \propto \exp(-p_1^2/\sigma)$ with $\sigma\approx 2500$. Dashed lines (in red) indicate the exponential localization $|\psi(p_1)|^2 \propto \exp(-|p_1|/\zeta)$ with $\zeta\approx 2.5$ and $7.7$ corresponding to $\lambda=0.1$ and $0.05$, respectively. The parameters are $K=5.0$, $\varepsilon=5$, and $\ehbar=0.06$.
 \label{QCDiffu}}
\end{center}
\end{figure}

Previous investigations on the Hermitian systems  (i.e., $\lambda=0$) have reported that the two-particle interaction is able to destroy the quantum coherence if the interaction is strong enough~\cite{rossi,bandy,adach,graha,park,Zhao09,Zhao10,Zhao20,Gadway,Rozenbaum17}.
As a consequence, the quantum diffusion is in consistence with its classical counterpart, that is the QCC. In the Hermitian case, the classical dynamics is governed by the mapping equation, with which we can numerically investigate the classical dynamics. We consider the case that the kick strength of the two particles are strong enough, i.e., $K=5$, so that the classical dynamics is fully chaotic. The QCC of Hermitian case is shown in Fig.~\ref{QCDiffu}(a), where one can see that both the classical and quantum mean energy of particle 1 increases linearly with time, i.e., $\langle p_1^2(t)\rangle = Dt$. In addition, the diffusion coefficient $D$ almost equals to that of single kicked rotor, i.e., $D\approx K^2/2$, since in our model the interaction strength is negligibly small in the semiclassical limits, i.e., $\ehbar \ll 1$ (see Eq.~\eqref{INTACTHamil}).
It is worth noting that the linear increase of the mean square of momentum, which is traditionally termed as normal diffusion, is a character of chaotic behavior.
We further investigate the momentum distribution for the appearance of QCC. Our results show that the momentum distribution is in the form of the Gaussian function $|\psi(p_1)|^2 \propto \exp(-p_1^2/\sigma)$, which is the brand mark of chaotic diffusion in momentum space and demonstrates the disappearance of quantum coherence due to the interaction (see Fig.~\ref{QCDiffu}(b) for $\lambda=0$).
%%%
%%%

Interestingly, for a specific value of $\lambda$ (e.g., $\lambda=0.05$ in Fig.~\ref{QCDiffu}(a)), the quantum diffusion of the system complies with the classically-chaotic diffusion during finite time interval, beyond which it gradually approaches to saturation. This clearly demonstrates the appearance of DL. From Fig.~\ref{QCDiffu}(a), one can also see that the saturation value of $\langle p_1^2\rangle$ decreases with the increase of $\lambda$, which demonstrates that the extent of DL grows as the
strength of the non-Hermitian driving increases~\cite{Zhao20b}. To confirm the appearance of DL, we numerically investigate the momentum distribution at the time when the mean value $\langle p_1^2\rangle$ saturates. Our results show that the quantum state is exponentially localized in momentum space, i.e., $|\psi(p_1)|^2 \propto \exp(-|p_1|/\zeta)$ (see Fig.~\ref{QCDiffu}(b) for $\lambda=0.05$), which is in a sharp contrast with the Gaussian distribution in the presence of QCC. Moreover, the localization length $\zeta$ decreases with the increase of $\lambda$ (see Fig.~\ref{QCDiffu}(b) for $\lambda=0.1$), which is consistent with the tendency of the mean energy $\langle p^2_1\rangle$ with the increase of $\lambda$, and thus is a clear evidence of DL.

\begin{figure*}[htbp]
\begin{center}
\includegraphics[width=17.0cm]{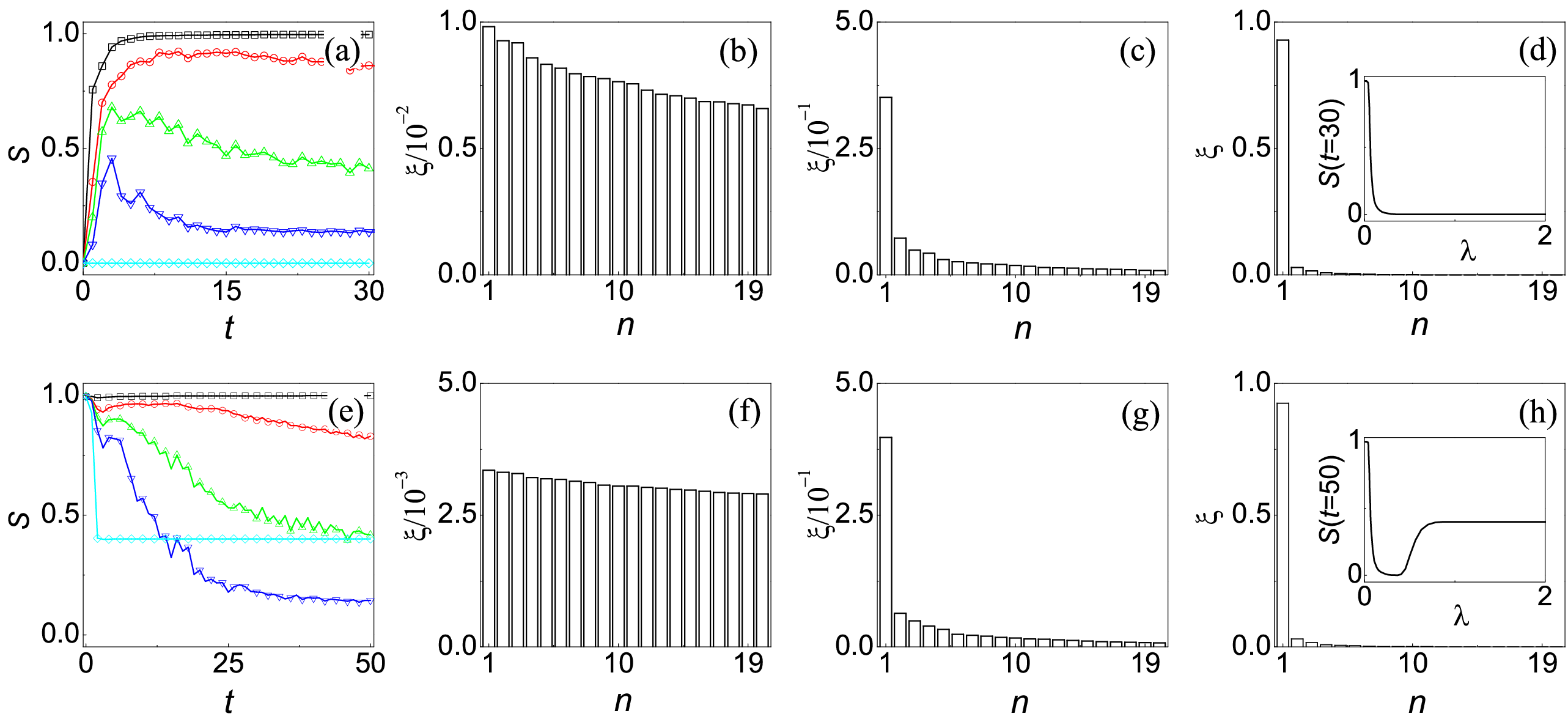}
\caption{(Color online) (a) Linear entropy $S$ versus time with $\lambda=0.0$ (squares), $0.05$ (circles), $0.07$ (up triangles), and $0.1$ (down triangles), and $1$ (diamonds). (b)-(d) The eigenvalue $\xi$ of the reduced density operator $\rho_{1}$ at the time  $t_n=30$ with $\lambda=0$ (b), $0.05$ (c), $0.1$ (d). Inset in (d): dependence of the $S$ at the time $t_n=30$ on $\lambda$. The initial state is a product state of the ground state $|\psi(t_0)\rangle = |00\rangle$.
Bottom panels same as in the above but for an initial state with the maximum entanglement, i.e., $|\psi(t_0)\rangle= \sum\psi_{n,n+1}|n,n+1\rangle$ with $\psi_{n,n+1}= \mathcal{N}_0 e^{- n^2/\sigma}$ and $\sigma=12000$.
In (e): $\lambda=0.0 $ (squares), $0.05$ (circles), $0.07$ (triangles), and $0.1$ (diamonds). In (f)-(h): the $\xi$ of $\rho_1$ at the time $t_n=50$ for $\lambda=0.0$ (f), $0.05$ (g), and $0.1$ (h). Inset in (h): the $S$ at the time $t_n=50$ versus $\lambda$.
Other parameters are the same as in Fig.~\ref{QCDiffu}.\label{entropy}}
\end{center}
\end{figure*}

It is known that during the process of QCC the entanglement between the subsystems grows. A commonly used quantity to measure the entanglement is the linear entropy
\begin{equation}
S(t)=1-\operatorname{Tr}[\rho_{1}^{2}(t)]\;.
\end{equation}
With the generation of entanglement, a pure state evolves to a mixed one, correspondingly the value of $S$ increases from zero to almost unity~\cite{Wang12,Wang20} (see Fig.~\ref{entropy}(a) for $\lambda=0$). This process is accompanied by the disappearance of quantum coherence. Therefore, it is believed that the decoherence effects induced by interaction result in the QCC of diffusion dynamics.
Our results show that, for a specific $\lambda$, the linear entropy rapidly increases to saturation as time evolves (see Fig.~\ref{entropy}(a) for $\lambda_2=0.05$). Interestingly, the saturation value decreases with the increase of $\lambda$, and is almost zero for sufficiently large $\lambda$ (see the inset in Fig.~\ref{entropy}(d)), which clearly demonstrates the reduce of entanglement by the non-Hermitian driving.
As a further step, we numerically investigate the eigenvalue $\xi$ of the reduced density operator $\rho_{1}$ at the time when the linear entropy saturates.
Our results show that for small $\lambda$ (e.g.,  $\lambda=0.05$ in Fig.~\ref{entropy}(b)), there are many eigenvalues. For intermediate value of $\lambda$ (e.g., $\lambda=0.07$ in Fig.~\ref{entropy}(c)), the number of $\xi$ with large values reduces. For large $\lambda$ (e.g., $\lambda=0.1$ Fig.~\ref{entropy}(d)), there is a $\xi$ whose value is almost unity, i.e., $\xi\approx 1$.
It is known that the appearance of many relatively large $\xi$ indicates that the time-evolved quantum state is a mixed state. For the appearance of a eigenvalue of unity, the quantum state becomes actually pure state. Therefore, our results demonstrate the transition of a quantum state from the mixed state to an almost pure state as  the strength of the non-Hermitian kicking potential increases. It is a strong evidence that the non-Hermitian driving effectively protects the quantum coherence.

We further consider the case that the initial state is of maximum entanglement, i.e., $|\psi(t_0)\rangle= \sum\psi_{n,n+1}|n,n+1\rangle$, for which the component is in the distribution of the Gaussian function $\psi_{n,n+1}= \mathcal{N}_0 e^{- n^2/\sigma}$ with $\mathcal{N}_0$ being the normalization constant and $\sigma=12000$. The value of linear entropy of this state is unity, i.e, $S(t_0)=1$. We find that for $\lambda=0$, the value of $S$ remains at unity, i.e., $S=1$, as time evolves. For small $\lambda$ (e.g., $\lambda=0.05$ in Fig.~\ref{entropy}(e)), the value of $S$ decays from unity as time evolves. For larger value of $\lambda$ (e.g., $\lambda=0.07$ and $0.1$ in Fig.~\ref{entropy}(e)), the value of $S$ rapidly decays from unity to saturation with time evolution.
We further investigate the saturation value of $S$ for different $\lambda$. Our numerical results show that, for $\lambda$ smaller than a threshold value $\lambda<\lambda_c$, the saturation value decreases from unity to almost zero with the increase of $\lambda$, and for $\lambda>\lambda_c$ it increases to a fixed value with an increasing $\lambda$ (see the inset in Fig.~\ref{entropy}(h)). It clearly demonstrates that the quantum coherence recovers with the increase of non-Hermitian driving. To confirm this finding, we numerically investigate the eigenvalue of the $\rho_1$ when the linear entropy saturates. Our numerical results show that, with the increase of the $\lambda$, the number of relatively large $\xi$ decreases, and for sufficiently large $\lambda$ there is a $\xi$ equal to almost unity (see Fig.~\ref{entropy}(f-h)). This again demonstrates the transition from a mixed state to pure state under the effects of non-Hermitian driving, which is a signature of the recovery of quantum coherence.
%%%%%

\section{Theoretical Analysis}\label{THEAnal}
\begin{figure}[htbp]
\begin{center}
\includegraphics[width=8.5cm]{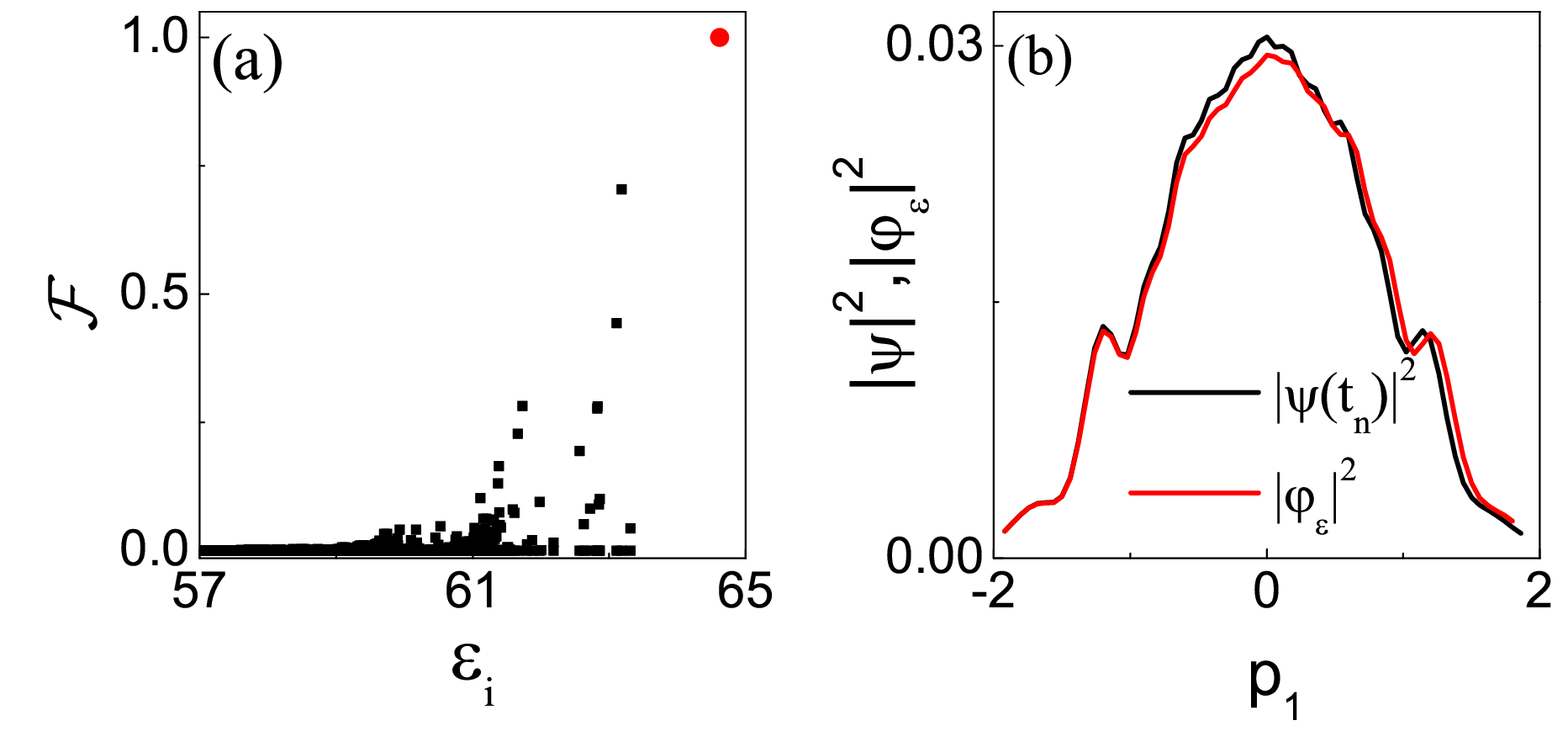}
\caption{(Color online) (a) Fidelity $\mathcal{F}=|\langle \varphi_{\varepsilon}|\psi(t_n)\rangle|^2$  between the quasieigenstates $|\varphi_{\varepsilon}\rangle$ and the quantum state $|\psi(t_n)\rangle$ with $t_n=100$. The red dot marks the value of $\mathcal{F}\approx1$ corresponding to maximum value of $\varepsilon_i=64.62$. (b) Comparison of momentum distributions between the quasieigenstate $|\varphi_{\varepsilon}\rangle$ with $\varepsilon_i=64.62$ (red line) and the quantum state $|\psi(t_n=100)\rangle$ (black line). The value of $\lambda$ is $\lambda=2.0$. Other parameters are the same as in Fig.~\ref{QCDiffu}. \label{fidelity}}
\end{center}
\end{figure}

The eigenequation of the Floquet operator has the expression $U|\varphi_{\varepsilon}\rangle = e^{-i\varepsilon}|\varphi_{\varepsilon}\rangle$, where $\varepsilon$ is the QEV and $|\varphi_{\varepsilon}\rangle$ is the corresponding QES. At the initial time, an arbitrary state can be expanded on the basis of the QESs, namely, $|\psi(t_0)\rangle = \sum_{\varepsilon} C_{\varepsilon}|\varphi_{\varepsilon}\rangle$. After the $n$th kick, the expansion of the state takes the form $|\psi(t_n)\rangle = U^{n}|\psi(t_0)\rangle =\sum_{\varepsilon} C_{\varepsilon}e^{-in\varepsilon}|\varphi_{\varepsilon}\rangle$.
It is worth noting that the QEVs become complex, i.e., $\varepsilon=\varepsilon_r + i\varepsilon_i$ in condition that the $\lambda$ is large enough. Accordingly, one can get $|\psi(t_n)\rangle = \sum_{\varepsilon} C_{\varepsilon}e^{n\varepsilon_i}e^{-in\varepsilon_r}|\varphi_{\varepsilon}\rangle$.
With time evolution, the components $C_{\varepsilon}$ with $\varepsilon_i >0$ will exponentially increase, and that of negative $\varepsilon_i$ exponentially decays. So, the quantum state $|\psi(t_n)\rangle$ will eventually evolve to the QES with the maximum $\varepsilon_i$. To confirm our analysis, we numerically investigate the fidelity between the QESs and the time-evolved state  $\mathcal{F}=|\langle \varphi_{\varepsilon}|\psi(t_n)\rangle|^2$ when DL appears while the linear entropy is almost zero for large enough $\lambda$ (e.g, $t_n=100$ and $\lambda=2.0$ in Fig.~\eqref{fidelity}(a)). Our result shows that the value of $\cal{F}$ is almost unity corresponding to the maximum value of $\varepsilon_i$, i.e., $\varepsilon_i^{max}=64.62$ (see Fig.~\eqref{fidelity}(a)), which demonstrates the coincidence of the quantum state with the QESs of $\varepsilon_i^{max}$. Note that the QESs with complex QEVs are not orthogonal, hence there are some nonzero values of $\mathcal{F}$.
We further display both the quantum state $|\psi(t_n)\rangle$ and the QES $|\varphi_{\varepsilon_i}\rangle$ of $\varepsilon_i^{max}$ in momentum space. The results again suggest that the two states are in perfect agreement with each other, both of which are extremely localized in momentum space (see Fig.~\eqref{fidelity}(b)).
Therefore, the appearance of DL is rooted in the localization of the QES with maximum $\varepsilon_i$. As the localization length of this QES is very small, it can be safely regraded as a pure state, which preserves the quantum coherence, hence the linear entropy is nearly zero (see the inset in Fig.~\ref{entropy}(e)). Even if we select a state with the maximum entanglement, it will rapidly evolve to the QES of maximum $\varepsilon_i$. This process corresponds to the decay of linear entropy from unity to a saturation level as shown in Fig.~\ref{entropy}(e). So the recovery of quantum coherence in the coupled bipartite system is due to the fact that the quantum states evolve to a QES with maximum values of imaginary QEV, moreover, this QES is virtually a pure state. Our finding establishes a new mechanism for the restoration of quantum coherence by non-Hermitian driving, which is not only significant in theoretical physics, but also has potential application in the fields of both quantum technology and quantum control.
%%%%%%%%%%%%%%%%%%%%%%%%%%%%%%%%%
%%%%%%%%%%%%%%%%%%%%%%%%%%%%%%%%%
%%%%%%%%%%%%%%%%%%%%%%%%%%%%%%%%%
%%%%%%%%%%%%%%%%%%%%%%%%%%%%%%%%%

\section{Summary}\label{ConcL}
In this work, we investigate the dynamics of the quantum diffusion and entanglement in a bipartite system with non-Hermitian kicking potential.
For a strong enough $\lambda$, the quantum diffusion of $\langle p_1^2 \rangle$ follows the classically-chaotic way during the short time interval, after which it saturates. The saturation values of $\langle p_1^2 \rangle$ decrease with the increase of $\lambda$, which clearly demonstrates the enhancement of DL by the non-Hermitian driving. The formation of DL corresponds to the reduce of entanglement between the two subsystems, for which the the saturation level of linear entropy decreases with the increase of $\lambda$. On the other hand, for maximum entanglement states, the linear entropy will decay from 1 to almost 0 during the evolving process, which again suggests the recovery of quantum coherence. As a consequence, the time-evolved state becomes a pure state, which is characterized by the formation of one significantly large eigenvalue of reduced density matrix. The mechanism is that a quantum state finally evolves to a QES whose imaginary part of the QEV is maximum. Our investigation reveals that non-Hermiticty is useful to protect quantum coherence, which provides new insights  in the fields of quantum information and quantum chaos.

\textcolor{blue}{\textit{Acknowledgements}---} We are grateful to Jie Liu, Hua Yan and Lewei He for stimulating discussions. W. Zhao. is supported by the National Natural Science Foundation of China (Grant Nos.12065009 and 11864014).
Z. Li. is supported by the Natural Science Foundation of China (Grant No. 11704132), the Science and Technology Program of Guangzhou No. 201905001.


\begin{thebibliography}{*}

\bibitem{Anderson58}
P. W. Anderson, Phys. Rev. {\bf 109}, 1492 (1958).\\ \textcolor{red}{DOI:10.1103/PhysRev.109.1492}

\bibitem{Condat87a}
C. A. Condat and T. R. Kirkpatrick, \prl {\bf 58}, 226 (1987). \\ \textcolor{red}{DOI:https://doi.org/10.1103/PhysRevLett.58.226}

\bibitem{Condat87b}
C. A. Condat and T. R. Kirkpatrick, \prb {\bf 36}, 6782 (1987).\\ \textcolor{red}{DOI:https://doi.org/10.1103/PhysRevB.36.6782}

\bibitem{Chabanov00}
A. A. Chabanov, M. Stoytchev, and  A. Z. Genack, Nature (London) {\bf 404}, 850 (2000). \\ \textcolor{red}{DOI:https://doi.org/10.1038/35009055}

\bibitem{Segev13}
M. Segev, Y. Silberberg, and D. N. Christodoulides,  Nature Photonics, {\bf 7}, 197 (2013). \\ \textcolor{red}{DOI:https://doi.org/10.1038/nphoton.2013.30}

\bibitem{Billy08}
J. Billy, V. Josse, Z. Zuo, A. Bernard, B. Hambrecht, P. Lugan, D. Cl\'ement, L. Sanchez-Palencia, P. Bouyer, and A.
Aspect, Nature (London) {\bf 453}, 891 (2008). \\ \textcolor{red}{DOI:https://doi.org/10.1038/nature07000}

\bibitem{Roati08}
 G. Roati, C. D'Errico, L. Fallani, M. Fattori, C. Fort, M. Zaccanti, G. Modugno, M. Modugno, and M. Inguscio, Nature (London) {\bf 453}, 895 (2008).\\ \textcolor{red}{DOI:https://doi.org/10.1038/nature07071}

\bibitem{Casati79}
G. Casati, B. V. Chirikov, F. M. Izrailev, and J. Ford, {\it in Stochastic
Behavior in Classical and Quantum Hamiltonian Systems},
edited by G. Casati and J. Ford, Lecture Notes in Physics
Vol. 93 (Springer, Berlin, 1979).

\bibitem{Izrailev90}
F. M. Izrailev, Phys. Rep. {\bf 196}, 299 (1990). \\ \textcolor{red}{DOI:https://doi.org/10.1016/0370-1573(90)90067-C}

\bibitem{Fishman82}
S. Fishman, D. R. Grempel, and R. E. Prange, \prl {\bf49}, 509 (1982). \\ \textcolor{red}{DOI:https://doi.org/10.1103/PhysRevLett.49.50}


\bibitem{Moore95}
F. L. Moore, J. C. Robinson, C. F. Bharucha, B. Sundaram, and M. G. Raizen, \prl {\bf 75}, 4598 (1995). \\ \textcolor{red}{DOI:https://doi.org/10.1103/physrevlett.75.4598}

\bibitem{Ammann98}
H. Ammann, R. Gray, I. Shvarchuck, and N. Christensen, \prl {\bf 80}, 4111 (1998). \\ \textcolor{red}{DOI:https://doi.org/10.1103/PhysRevLett.80.4111}

\bibitem{dArcy01}
M. B. d\'Arcy, R.M.Godun, M. K. Oberthaler,D. Cassettari, and G. S. Summy, \prl {\bf 87}, 074102 (2001). \\ \textcolor{red}{DOI:https://doi.org/10.1103/PhysRevLett.87.074102}

\bibitem{Ryu06}
C. Ryu, M. F. Andersen, A. Vaziri, M. B. d'Arcy, J. M. Grossman, K. Helmerson, and W. D. Phillips, \prl {\bf 96}, 160403 (2006). \\ \textcolor{red}{DOI:https://doi.org/10.1103/PhysRevLett.96.160403}

\bibitem{Talukdar10}
I. Talukdar, R. Shrestha, and G. S. Summy, \prl {\bf 105}, 054103 (2010). \\ \textcolor{red}{DOI:https://doi.org/10.1103/PhysRevLett.105.054103}

\bibitem{Lopez12}
M. Lopez, J. F. Clement, P. Szriftgiser, J. C. Garreau, and D. Delande, \prl {\bf 108}, 095701 (2012).\\ \textcolor{red}{DOI:https://doi.org/10.1103/physrevlett.108.095701}

\bibitem{Gadway13}
B. Gadway, J. Reeves, L. Krinner, and D. Schneble, \prl {\bf 110}, 190401 (2013). \\ \textcolor{red}{DOI:https://doi.org/10.1103/PhysRevLett.110.190401}
\bibitem{Zhaohong20}
Z. Wang, W. Fu, Y. Zhang, and H. Zhao, \prl{\bf 124} 186401 (2020).\\ \textcolor{red}{DOI:https://doi.org/10.1103/PhysRevLett.124.186401}


\bibitem{Fishman12}
S. Fishman, Y. Krivolapov, and A. Soffer, Nonlinearity {\bf 25}, R53 (2012).\\ \textcolor{red}{DOI:https://doi.org/10.1088/0951-7715/25/4/R53}

\bibitem{Flach13}
Ch. Skokos, I. Gkolias, and S. Flach, \prl {\bf 111}, 064101 (2013). \\ \textcolor{red}{DOI:https://doi.org/10.1103/PhysRevLett.111.064101}

\bibitem{Veksler09}
H. Veksler, Y. Krivolapov and S. Fishman, \pre {\bf 80}, 037201 (2009). \\ \textcolor{red}{DOI:https://doi.org/10.1103/physreve.80.037201}

\bibitem{Pikovsky11}
A. Pikovsky and S. Fishman, \pre {\bf 83}, 025201 (2011). \\ \textcolor{red}{DOI:https://doi.org/10.1103/PhysRevE.83.025201}

\bibitem{Michaely12}
E. Michaely and S. Fishman, \pre {\bf 85}, 046218 (2012). \\ \textcolor{red}{DOI:https://doi.org/10.1103/PhysRevE.85.046218}

\bibitem{Vakulchyk18}
I. Vakulchyk, M. V. Fistul, and S. Flach, \prl {\bf 122}, 040501 (2019). \\ \textcolor{red}{DOI:https://doi.org/10.1103/PhysRevLett.122.040501}

\bibitem{Cherroret14}
N. Cherroret, B. Vermersch, J. C. Garreau, and D. Delande, \prl {\bf 112}, 170603 (2014). \\ \textcolor{red}{DOI:https://doi.org/10.1103/PhysRevLett.112.170603}


\bibitem{Gligoric13}
G. Gli\'gori\'c, K. Rayanov, and S. Flach, EPL, {\bf 101}, 10011 (2013). \\ \textcolor{red}{DOI:https://doi.org/10.1209/0295-5075/101/10011}

\bibitem{Mieck05}
B. Mieck and R. Graham, J. Phys. A, {\bf 38}, L139 (2005). \\ \textcolor{red}{DOI:https://doi.org/10.1088/0305-4470/38/7/L05}

\bibitem{Guarneri17}
I. Guarneri, \pre {\bf 95}, 032206 (2017). \\ \textcolor{red}{DOI:https://doi.org/10.1103/PhysRevE.95.032206}


\bibitem{Zhao16}
W. Zhao, J. Gong, W. Wang, G. Casati, J. Liu, and L. Fu, \pra {\bf 94}, 053631 (2016). \\ \textcolor{red}{DOI:https://doi.org/10.1103/PhysRevA.94.053631}

\bibitem{Zhao19JPA}
W. Zhao, J. Z. Wang and W. Wang, J. Phys. A: Math. Theor. {\bf 52} 305101 (2019). \\ \textcolor{red}{DOI:https://doi.org/10.1088/1751-8121/ab1cde}


\bibitem{Rylands20}
C. Rylands, E. B. Rozenbaum, V. Galitski, and R. Konik, \prl {\bf 124}, 155302 (2020). \\ \textcolor{red}{DOI:https://doi.org/10.1103/PhysRevLett.124.155302}



\bibitem{rossi} D. Rossini, G. Benenti, and G. Casati, \pre {\bf 74}, 036209 (2006). \\ \textcolor{red}{DOI:https://doi.org/10.1103/PhysRevE.74.036209}

\bibitem{bandy} J. N. Bandyopadhyay, Europhys. Lett. {\bf 85}, 50006 (2009).\\
\textcolor{red}{https://doi.org/10.1209/0295-5075/85/50006}

\bibitem{adach}
S. Adachi, M. Toda, and K. Ikeda, \prl {\bf 61}, 659 (1988).\\
\textcolor{red}{DOI:https://doi.org/10.1103/PhysRevLett.61.659}

\bibitem{graha} R. Graham and A. R. Kolovsky, Phys. Lett. A. {\bf 222}, 47 (1996).\\ \textcolor{red}{DOI:https://doi.org/10.1016/0375-9601(96)00623-8}

\bibitem{park} H. K. Park and S. W. Kim, \pra {\bf 67}, 060102(R) (2003). \\ \textcolor{red}{DOI:https://doi.org/10.1103/PhysRevA.67.060102}


\bibitem{Zhao09}
W. L. Zhao and Q. L. Jie, Commun. Theor. Phys. {\bf 51}, 465 (2009). \\ \textcolor{red}{DOI:https://doi.org/10.1088/0253-6102/51/3/17}

\bibitem{Zhao10}
W. L. Zhao, Q. L. Jie, and B. Zhou, Commun. Theor. Phys. {\bf 54}, 247 (2010).\\ \textcolor{red}{DOI:https://doi.org/10.1088/0253-6102/54/2/09}

\bibitem{Zhao20}
W. L. Zhao and Q. L. Jie, Chin. Phys. B {\bf 29},   080302 (2020). \\ \textcolor{red}{DOI:https://doi.org/10.1088/1674-1056/ab8a3a}

\bibitem{Gadway}
B. Gadway, J. Reeves, L. Krinner, and D. Schneble, \prl {\bf 110}, 190401 (2013).\\  \textcolor{red}{DOI:https://doi.org/10.1103/PhysRevLett.110.190401}

\bibitem{Rozenbaum17}
E B Rozenbaum  and V  Galitski, \prb {\bf 95} 064303 (2017). \\ \textcolor{red}{DOI:https://doi.org/10.1103/PhysRevB.95.064303}

\bibitem{Ashida20}
Y. Ashida, Z. Gong, and M. Ueda, arxiv: 2006.01837 (2020).

\bibitem{Zhao20C}
W. L. Zhao, L. W. Zhou, J. Liu, P. Q. Tong,  and K. Q. Huang, \pra {\bf 102}, 062213 (2020). \\ \textcolor{red}{DOI:https://doi.org/10.1103/PhysRevA.102.062213}

\bibitem{Zhao20B}
W. L. Zhao,  P. K. Gong, J. Z. Wang, and Q. Wang,  Chinese Phys. B. {\bf 29}, 120302 (2020). \\ \textcolor{red}{DOI:https://doi.org/10.1088/1674-1056/abc0dc}

\bibitem{Suchkov16}
S. V. Suchkov, A. A. Sukhorukov, J. H. Huang, S. V. Dmitriev, C. H. Lee, and Y. S. Kivshar, Laser Photonics Rev. {\bf 10}, 177 (2016). \\ \textcolor{red}{DOI:https://doi.org/10.1002/lpor.201500227}

\bibitem{Liying19}
Y. Li. {\it et al.,} Science, {\bf 364}, 170  (2019).\\ \textcolor{red}{DOI:10.1126/science.aaw6259}


\bibitem{Moiseyev2011}
N. Moiseyev, {\it Non-Hermitian quantum mechanics}, (Cambridge University Press, Cambridge, UK, 2011).

\bibitem{Zhao20b}
K. Q. Huang, J. Z. Wang, W. L. Zhao, and Jie
Liu, J. Phys.: Condens. Matter. {\bf 33}, 055402 (2021). \\ \textcolor{red}{DOI:https://doi.org/10.1088/1361-648X/abbcf8}


\bibitem{Wang12}
W. G. Wang, L. He, and J. Gong, \prl {\bf 108} 070403 (2012). \\ \textcolor{red}{DOI:https://doi.org/10.1103/PhysRevLett.108.070403}


\bibitem{Wang20}
W. G. Wang, \pre {\bf 102}, 012127 (2020). \\ \textcolor{red}{DOI:https://doi.org/10.1103/PhysRevE.102.012127}

\end{thebibliography}
\end{document}